\documentclass[12pt,preprint]{aastex}

\newcommand{\halpha}{H$\alpha$}
\newcommand{\hii}{H\thinspace II}

\shorttitle{Dwarf Galaxies in Abell 1367}
\shortauthors{Sakai et al.}

\begin{document}

\title{Discovery of a Group of Star-Forming
Dwarf Galaxies in Abell~1367}

\medskip
\author{Shoko Sakai\altaffilmark{1}}
\affil{Department of Astronomy, University of California, Los Angeles,
Los Angeles, CA, 90095-1562}
\email{shoko@astro.ucla.edu}

\author{Robert C.\ Kennicutt, Jr.\altaffilmark{1,2}}
\affil{Steward Observatory, University of Arizona, Tucson, AZ  85721}
\email{rkennicutt@as.arizona.edu}

\author{J.\ M.\ van der Hulst}
\affil{Kapteyn Astronomical Institute, Postbus 800, Groningen, NL 9700 AV, 
Netherlands}
\email{vdhulst@astro.rug.nl}

\and

\author{Chris Moss\altaffilmark{1}}
\affil{Liverpool John Moore's University, Astrophysics Research Institute,
Birkenhead CH41 1LD, United Kingdom}
\email{cmm@astro.livjm.ac.uk}

\altaffiltext{1}{Visiting Astronomer, Kitt Peak National Observatory.
KPNO is operated by AURA, Inc.\ under contract to the National Science
Foundation.}

\altaffiltext{2}{Part of the observations reported here were obtained at 
the MMT Observatory, a joint facility of the
University of Arizona and the Smithsonian Institution.} 

\begin{abstract}

We describe the properties of a remarkable group of actively star-forming
dwarf galaxies and \hii\ galaxies in the Abell~1367 cluster, which were
discovered in a large-scale \halpha\ imaging survey of the cluster.
Approximately 30 \halpha-emitting knots were identified in a region
approximately 150 kpc across, in the vicinity of the spiral galaxies
NGC 3860, CGCG 97-125 and CGCG 97-114.  Follow-up 
imaging and spectroscopy reveals that some of the knots are
associated with previously uncataloged dwarf galaxies (M$_B$ = $-15.8$ to $-16.5$), 
while others appear to be isolated \hii\ galaxies or intergalactic
\hii\ regions.  
Radial velocities obtained for several of the knots show that
they are physically associated with a small group or subcluster
including CGCG~97-114 and CGCG~97-125. 
No comparable concentration of emission-line
objects has been found elsewhere in any of the eight
northern Abell clusters surveyed to date.  
The strong \halpha\ emission in the objects and their high spatial
density argue against this being a group of normal, unperturbed dwarf
galaxies.  Emission-line spectra of several of the knots also show
some to be anomalously metal-rich relative to their luminosities.
The results suggest that many of these objects were formed or triggered by 
tidal interactions or mergers involving CGCG 97-125 and other members of 
the group.  A Westerbork Synthesis Radio Telescope HI map of the region
shows direct evidence for tidal interactions in the group.
These objects may be related to the tidal dwarf galaxies found in
some interacting galaxy pairs, merger remnants, and compact groups. 
They could also represent evolutionary precursors to the class of isolated
ultracompact dwarf galaxies that have been identified in the Fornax cluster.

\end{abstract}

\keywords{galaxies: dwarf -- galaxies: formation -- galaxies: evolution --
galaxies: clusters: individual (Abell~1367)}

\section{INTRODUCTION}

Until recently most surveys for star-forming galaxies in the nearby
universe have been restricted to imaging of previously cataloged
objects, or wider-field prism surveys that are mainly sensitive to
strong emission-line galaxies.  These have provided a relatively 
complete inventory of massive galaxies and starburst galaxies,
but they provide much less complete information on the population of
star-forming dwarf galaxies.  Large numbers of very nearby star-forming
dwarfs have been studied (e.g., Terlevich et al. 1991, 
Hunter, Hawley, \& Gallagher 1993, van Zee 2000, 2001), but complete
star formation inventories, extending across the full range of galaxy
types {\it and} masses, are lacking.

As part of a larger effort to obtain complete inventories of star
formation rates (SFRs) in nearby galaxy samples, we have carried 
out a deep, wide-field H$\alpha$ survey of
nearby clusters of galaxies using the MOSAIC CCD camera on the 0.9 m telescope
at Kitt Peak National Observatory (Sakai et al. 2001a).
Each field covers one square degree, and we have obtained high-quality data
for 25 fields in 8 nearby northern Abell clusters, in the radial velocity
range 3000 $-$ 8000 km~s$^{-1}$.  
A total of six fields were observed in Abell~1367, and $\sim$250 
H$\alpha$-emitting galaxies were detected (Sakai et al. 2002, in preparation). 
During the course of this analysis we discovered
an unusual concentration of \halpha-emitting dwarf galaxies
and \hii\ galaxies in the central field of the Abell 1367 cluster.
A brief report was given in Sakai et al. (2001b).  
Recently Iglesias-Paramo et al. (2002) presented the results of an independent
H$\alpha$ survey of the center of A1367, and they comment specifically
on the unusual properties of this region of the cluster (see the appendix
of their paper).  
Our follow-up spectroscopy (\S 4) shows that these objects are part of a low
velocity dispersion group or subcluster within or behind the main
cluster, which also contains two Zwicky galaxies: CGCG~97-114 and CGCG~97-125.

Although the discovery of emission-line dwarf galaxies in Abell~1367  
is not extraordinary in itself, the concentration of such objects
in this region is very unusual --- to date no other such concentrations
have been found in any of the eight clusters we surveyed.  Moreover,  
emission-line spectra obtained for several of the knots reveal chemical
properties that are inconsistent with the scenario that this region is
simply a grouping of normal star-forming dwarf galaxies.  Instead, the observational
evidence suggests that at least some of the objects
are the products of galaxy interactions or other environmental processes 
within the group, or in conjunction with the larger A1367 cluster. Consequently
this serendipitously discovered group may offer valuable clues to the 
physical processes that influence the evolution and formation of dwarf
galaxies in groups and clusters.  

The remainder of the paper is organized as follows.  
In \S2 we discuss the data collected on this region, including 
the \halpha\  and follow-up broadband imaging, emission-line
spectroscopy, and HI aperture synthesis observations.  In \S3 we
use these data to characterize the nature of the galaxy group and
measure the SFRs and basic physical properties of the star-forming
dwarfs.  Finally in \S4 we consider possible physical explanations
for the nature and formation of these objects, and tentatively conclude
that they are a combination of pre-existing dwarf galaxies, intergalactic
HII regions, and possibly newly formed dwarf galaxies, all triggered
by tidal interactions between the larger members
of the group.  We also place these results in the context of other
discussions of tidally-formed dwarf galaxies (e.g., Mirabel, Dottori,
\& Lutz 1992), and \hii\ regions (Iglesias-Paramo \& V\'ilchez 2001),
and the recent discovery of compact blue galaxies
in the Fornax cluster (Drinkwater et al. 2001, Phillips et al. 2001).

\section{DATA}

Here we briefly describe the wide-field MOSAIC \halpha\
imaging in which the objects were discovered, and which have been used
to quantify the star formation properties of the group members.
In addition we describe follow-up optical imaging, spectroscopy,
and HI observations, which were obtained in an effort to understand
the physical nature and formation of the dwarfs.

\subsection{MOSAIC Imaging}

The \halpha\ images were obtained under photometric 
conditions in 1999 Feb, using the MOSAIC-1 camera on the KPNO 0.9m telescope.
The camera is comprised of eight 2048$\times$4096 SITe CCDs with 
15$\mu$ pixels, corresponding to 0.43$^{\prime\prime}$/pixel, and a 
total aereal coverage of $59^{\prime} \times 59{^\prime}$ on the sky.
The observation of the central field of Abell~1367 was centered at
($11^h44^m09^s$,$19^{\circ}52^{\prime}33^{\prime\prime}$) (J2000).
Each observation consisted of five dithered 900 s exposures, using
three filters centered at 6615 \AA\ (Ha4), 6695 \AA\ (Ha12), and 
6736 \AA\ (Ha16).  
The effective bandwidth of each filter is approximately 80 \AA.
Since the mean velocity and velocity dispersion of Abell~1367 are
6595 km~s$^{-1}$ and 879 km~s$^{-1}$ respectively (Struble \& Rood 2001),
two on-band filters (Ha12, Ha16) were used for the detection
of H$\alpha$ emission, and Ha4 for the continuum observation.
For most galaxies the filters admitted the adjacent [NII] lines
at 6548 \AA\ and 6583 \AA\ as well, and for the purposes of this
paper we will use ``\halpha" to refer to the combined \halpha$+$[NII] 
emission.

The raw images were processed within the NOAO IRAF package\footnote
{IRAF is distributed by the National Optical Astronomy 
Observatories, which are operated by the Association of Universities for 
Research in Astronomy, Inc., under cooperative agreement with the National
Science Foundation.},
using standard procedures as documented
in the NOAO MOSAIC manual (available on-line at URL:
http://www.noao.edu/kpno/mosaic).  The flatfield correction for each filter was
made by taking a median of five dome flats, which was sufficient for
these narrowband images.  Two extra steps
needed to be performed with the reduction of the MOSAIC images, compared
to conventional single-detector cameras, in order to correct for the
variable pixel scale, which changes by 6\% from the center to the edge of
the field.  The dithered images were reinterpolated to a tangent-plane 
projection, to provide a constant angular scale.  The typical image
quality on the combined and regridded images, set by seeing and the 
reinterpolation process, is approximately 1.3\arcsec\ FWHM.

The flux zeropoints for the emission-line images were calibrated
using observations of spectrophotometric standards from Massey et al.
(1988), and analyzed using standard techniques for narrowband images
(e.g., Kennicutt \& Kent 1983).  The 75 min combined exposures 
provide 5-$\sigma$ emission-line detections of extended sources to 
approximately $10^{-15}$ ergs~cm$^{-2}$~s$^{-1}$, which corresponds to
an absolute luminosity of $\sim10^{39}$ ergs~s$^{-1}$, or a total SFR 
of $\sim$0.01 M$_\odot$~yr$^{-1}$, using the calibration
of Kennicutt (1983).  The detection limits for single, compact
\hii\ regions are approximately an order of magnitude fainter.

\subsection{Deep Imaging}

Most of the faint objects discovered in the \halpha\ images have
not been cataloged, and for many of them no continuum counterparts
were visible on the MOSAIC camera red continuum images described below.
In order to identify any faint parent galaxies associated with any of
the objects, and measure their basic photometric properties, we obtained
follow-up deep $B$ and $R$ imaging in February 2001.  The data were obtained using
the 2KCCD imager on the Bok 2.3 m telescope at Steward Observatory.  The camera
uses a 2048$\times$2048 thinned Loral CCD detector, which was binned
3$\times$3 to yield 0\farcs45 pixels and a field of view of 205 arcsec.
A series of exposures were obtained in each filter, providing 
combined exposure times of 3600s in $B$ and 1800s in $R$.  Data
were taken under photometric conditions, with seeing varying between
1.1\arcsec\ $-$ 1.4\arcsec\ FWHM.

We also used the same camera in March 2002 to obtain a series of 
deep \halpha\ images of the region.  The 2KCCD imager was used with
a 75 \AA\ FWHM filter centered at 6730 \AA\ for the on-band image,
with an accompanying $R$ image used for continuum subtraction.  Three
exposures totaling 3600 sec in exposure time were obtained, yielding
images about 5 times deeper (in effective exposure time) than the
MOSAIC images.

The images were reduced following standard CCD reduction procedures;
they were bias-subtracted and flatfielded using IRAF routines.
For the broadband imaging standard stars in the RU~149 field (Landolt 1992) 
were observed several 
times during the night and were used to calibrate the photometric zero points.
The \halpha\ images were reduced using the same procedures as described
in the previous section.

\subsection{Emission-Line Spectroscopy}

Before we could interpret the images it was important to obtain
radial velocities for some of the emission knots, in order to confirm whether
they were associated with Abell 1367 and to place constraints
on the kinematic properties of the system.  Exploratory spectra were obtained
for several of the objects on four nights between April 1999 and April 2000,
using the B\&C CCD Spectrograph on the Steward Observatory 2.3 m Bok telescope.
Although the immediate objectives were to determine radial velocities
for the knots, the spectra were of sufficient quality to impose some
constraints on the excitation and abundance properties of the regions
as well, as discussed in the next section.

The B\&C spectrograph was used with a 400 gpm grating and a 
1200 $\times$ 800 thinned Loral CCD detector, which
provided coverage of the spectral region 3650 -- 6950 \AA.
A slit width of 2\farcs5 yielded a resolution of 7.5 \AA.
Observations were carried out with the galaxies near the
zenith to eliminate any problems of atmospheric dispersion.
Standard calibration and long-slit reduction procedures
were employed, using standard stars from Massey et al. (1988)
and the TWODSPEC programs in IRAF.
Wavelengths of telluric night-sky lines in the spectra were used
to apply a second-order zeropoint correction to the radial velocity scales.

The target \hii\ regions were quite faint
($f(H\alpha) \sim 0.7 - 4 \times 10^{-15}$ ergs cm$^{-2}$ s$^{-1}$),
requiring blind offsets from nearby stars and bright
galaxies, as derived from the MOSAIC images.  Typical
integration times were 900 -- 2400 sec, sufficient to
confirm the redshifts of the objects, and in some cases 
derive approximate line ratios for the strongest features
(H$\alpha$, H$\beta$, [OII], [OIII], [NII]).  H$\alpha$
was easily the best detected line, and it was used
to measure the radial velocities of the knots, with a
typical measurement uncertainty of $\pm$50 km s$^{-1}$.
Usable spectra were obtained for seven objects 
(knots Dw1-a,b; Dw3-a,c; K2-a,b; CGCG 97-125-b in Figure~4).
We also observed two \hii\ regions in the main disk of the 
disrupted spiral galaxy CGCG 97-125,
to provide a reference and to test the hypothesis that some
of the intergalactic regions may have originated from gas
removed from the larger galaxies.

In January 2002, spectra for four \hii\ regions in the newly identified
dwarf galaxy Dw 1 (Table~2, Figure~4) and in CGCG 97-125 were
observed using the Blue Channel spectrograph on the newly refurbished
6.5 m MMT telescope.  The spectrograph was equipped with a 500 gpm
grating blazed at 5410 \AA, and a 3072 $\times$ 1024 thinned
Loral CCD detector, which provided spectra of the region 3600 $-$
7000 \AA\ with a resolution of 5 \AA.  Pairs of \hii\ regions in Dw~1 and
CGCG~97-125 were observed in 
single pointings for a total exposure time of 2400 seconds.
These yielded high S/N detections of the principle diagnostic 
emission-lines, and an independent check on the accuracy of the
spectra obtained at the 2.3 m telescope.  The data were reduced using
the same procedures as described above.

\subsection{HI Aperture Synthesis Imaging}

In order to place further constraints on the origin of the 
dwarf galaxies and emission knots, we obtained aperture synthesis HI data 
using the Westerbork Synthesis Radio Telescope (WSRT).
The WSRT observations were centered around NGC~3860, CGCG 97-114 and 
CGCG 97-125. The
detailed observing parameters are given in Table~1. The velocity range
covers the velocities of CGCG 97-114 and CGCG 97-125, but not the
lower velocity of NGC~3860 (5595 km s$^{-1}$, see next section).

We used the Miriad package (Sault et al. 1995) for the editing,
calibration and Fourier inversion of the data, and did the subsequent
analysis in Gipsy (van der Hulst et al. 1992, Vogelaar and Terlouw
2001). The WSRT is now equipped with a new correlator and cooled
21-cm front-ends on all telescopes, so that it becomes worthwhile to
use all possible telescope combinations for imaging the field of view
and increase significantly the sensitivity. The final resolution of
the data is a compromise between optimal sensitivity for detecting low
surface brightness HI, and retaining sufficient resolution to be able
to resolve the galaxies and their immediate environs.

The r.m.s. noise in the channel maps is 0.4 mJy which for the 
$18^{\prime\prime}\times 50^{\prime\prime}$ resolution corresponds to
a brightness temperature of 0.25 K. The corresponding 3$\sigma$ detection 
limit in column density is $2.4 \times 10^{19}$ cm$^{-2}$. The limiting mass
for a 50 km s$^{-1}$ wide profile is $1.9 \times 10^8$ M$_{\odot}$.

\section{RESULTS}

The structure of the CGCG 97-114/125 region is illustrated in Figure 1, which
shows deep $B$ and $R$ images of the field.  In Figure 2, we
show the combined \halpha+[NII] image.
The most prominent galaxy in the field is NGC~3860, but the radial
velocity of this galaxy is offset from most of the objects in the
field by 2600 km s$^{-1}$ (5600 km s$^{-1}$ for NGC~3860
vs 8200 $-$ 8300 km s$^{-1}$ from the main group (de Vaucouleurs et al.
1991, Giovanelli et al. 1997, Haynes et al. 1997), 
so it probably lies in the foreground to the group.  We will refer to
this region as the CGCG~97-114/125 Group for convenience,
based on the real physical association of these objects.

\subsection{Properties of the CGCG~97-114/125 Group}

The center of this group is located at ($11^h44^m50.6^s$,
$19^{\circ}46^{\prime}56^{\prime
\prime}$)(J2000).  This lies at a projected distance of only 6\arcmin\ 
(0.2 Mpc) southeast of the center of the A1367 cluster (Abell, Corwin,
\& Olowin 1989).  The location of the group is shown in Figure 3, which  
plots the spatial and velocity distributions of the galaxies in the 
clusters, as taken from the NASA/IPAC Extragalactic Database 
(NED)\footnote{This research has made use of the NASA/IPAC
Extragalactic Database (NED) which is operated by the Jet Propulsion 
Laboratory, California Institute of Technology, under contract with 
the National Aeronautics and Space Administration.}.  
The approximate location of the brighter members
of the CGCG~97-114/125 group is 
indicated by the open circle in each plot.  These plots show that the group
lies at the upper edge of the velocity distribution, displaced by
approximately 1700 km s$^{-1}$ from the centroid of A1367 as a whole,
implying that either the group is 
infalling into the central core of the A1367 cluster from the foreground,
or that it lies in the background to the cluster, as an extended part
of the Coma-A1367 supercluster.  The group spans approximately 4 $-$ 5
arcmin on the sky, which corresponds to a linear diameter of 110 $-$
130 kpc, if we adopt the distance to A1367 of 93 Mpc from Sakai et al.
(2000).  If the group lies in the background at its Hubble flow distance 
(110 kpc for H$_0$ = 75 km s$^{-1}$ Mpc$^{-1}$) the linear size would
increase to 130 $-$ 165 kpc.

\subsection{Identification of Dwarf and HII Galaxies}

Some of the emission-line galaxies and knots in this field were 
immediately apparent
on the raw \halpha\ images, even before continuum removal.  To obtain
a more complete data set, continuum-subtracted \halpha\ line images 
were obtained by shifting and scaling the Ha4 continuum images to
the same scales and subtracting them from the on-band Ha16 and Ha12
images.  These subtracted images were then blinked with
the continuum images to identify the objects visually.  
All of the emission-line objects were visible only on the Ha16 image,
but not on the Ha12 image (with the exception of NGC~3860).
Several other faint emission-line structures were identified later on the
deeper 2.3 m telescope \halpha\ images.  The
objects found are indicated in Figure 4.

The galaxies and objects seen in this group fall quite cleanly into three
classes: (1) previously cataloged Zwicky galaxies with multiple \hii\ regions
in their disks; (2) previously unidentified dwarf galaxies, identified in
Figure~4 as $Dw1$, $Dw2$ and $Dw3$;  (3) isolated intergalactic \hii\
knots, with no detectable underlying stellar components (e.g., 
$K1$ and $K2$ in Figure~4).  Table~2 summarizes the properties of
the galaxies and Table 3 lists positions and properties of the isolated
\hii\ knots.

\subsubsection{Zwicky Galaxies}

As described earlier, the group is defined by the galaxy 
pair CGCG~97-114 (also known as NGC 3860B) and CGCG 97-125,
and possibly two other cataloged 
objects, CGCG~97-113 and A1367[BO85] 121 (Butcher \& Oemler 1985).
Neither of the latter two objects have published radial velocities
(NED lists a velocity of 6413 km s$^{-1}$ for CGCG~97-113 but
no source is given), so their physical association with this group
is uncertain.  

CGCG~97-114 and 97-125 are luminous galaxies (M$_B \sim -19$ to $-19.5$)
classified in NED as irregular and spiral, 
respectively, and both show strong \halpha\ emission, with 
\halpha+[NII] equivalent widths of 44 \AA\ and 29 \AA\, respectively.
These values are typical for normal late-type spiral and irregular galaxies
(Kennicutt \& Kent 1983).  
Both are resolved into bright \hii\ regions with luminosities
(above our detection limit) of order $10^{38} - 10^{39}$ ergs s$^{-1}$,
comparable to giant \hii\ regions in the Local group and again quite
typical for late-type spiral and irregular galaxies of these masses.
Although most of these \hii\ regions lie well within the optical
confines of their parent galaxies (e.g., 
CGCG~97-114$c$, CGCG~97-125$d$ and $c$ in Figure 4), several bright
and very compact knots lie well outside the optical disks 
(e.g., knots $a$ and $b$ in both galaxies), and it is not clear whether
they are bound to parent galaxies or are separate objects (or satellites).

CGCG~97-125 and CGCG~97-114 both show peculiar morphologies in the 
\halpha\ images, as is best seen in Figure 2.  The broadband images 
of CGCG~97-125 show the presence of a  
faint, diffuse shell-like feature, which is characteristic of dynamically
disturbed galaxies or merger remnants (e.g., Malin \& Carter 1983,
Schweizer \& Seitzer 1988).  The broadband and \halpha\
images reveal a faint tail or bridge-like feature extending to the west
of CGCG~97-125, along with a faint arm of \hii\ knots (K2 in Fig.\ 2)
between the two galaxies.  These hint at a possible tidal bridge, but
the detected features are too faint and discontinuous to indicate
this for certain.  Neither of the other cataloged galaxies 
(CGCG~97-113, A1367[BO85]121) show detectable \halpha\ emission,
apart from possible low-level nuclear emission.
Finally, some of the HII regions in the vicinity of Zwicky galaxies, such as
CGCG~97-114$a$ and $b$ appear almost isolated, not physically associated
with the galaxy; there is no obvious underlying stellar distribution, even on
our deep broadband image.  

\subsubsection{Dwarf Galaxies}

The three dwarf galaxies Dw1, Dw2, and Dw3 (Table~2, Figure~4)
were virtually invisible on
the original continuum images taken using the MOSAIC camera, but
they were clearly identified in our follow-up deep broadband imaging.
All of them display highly irregular and possibly disturbed structures.

$Dw1$ consists of several luminous \hii\ regions in the original 
MOSAIC images, but the broadband images show that these reside within
a somewhat more diffuse structure of blue stars.  The $B$-band magnitude
of 18.5 corresponds to an absolute magnitude $M_B = -16.5$ for an
assumed distance modulus of 34.84 mag (Sakai et al. 2000).  This is
roughly comparable to the Small Magellanic Cloud, for example, 
but the stellar structure is less concentrated 
and more chaotic in appearance.  The galaxy also is unusually blue, with
$B-R = 0.05 \pm 0.07$ mag.  

Dw2 has an even more peculiar morphology.  In the deep $B$-band
image (Figure 2 left panel), Dw2 shows a V-shaped diffuse underlying
stellar distribution, with one very bright concentration which also shows
H$\alpha$-emission.  It could be tidally connected to CGCG 97-125
(and possibly to Dw3), but the proximity to NGC 3860 is 
mostly likely a superposition (recall that the radial velocity of the latter is 
2600 km~s$^{-1}$ lower).  The color of Dw2 is also blue, with
$B - R = 0.53 \pm 0.14$ (Table~2).

Dw3 is the faintest of the three newly identified dwarf irregular
galaxies ($M_B = -15.8$).  Its structure is similar to that of Dw1, with
a diffuse distribution of stars connecting the younger \hii\ regions,
and no evidence of a central concentration in $B$ or $R$.  It again is 
extremely blue, with $B-R=0.20 \pm 0.21$.

\subsubsection{Isolated \hii\ Galaxies and \hii\ Regions}

Over a dozen of the emission regions in the group do not show
any obvious association with any of galaxies, either the luminous
CGCG galaxies or the fainter dwarfs described above.  It is conceivable
that one or two could be very distant background galaxies with 
[OIII]$\lambda$5007 or [OII]$\lambda$3727 lines redshifted into
the \halpha\ bandpasses, but all of the knots measured spectroscopically
to date have radial velocities placing them in this group (\S 3.4).
Several of the knots may lie in the outer extended disks or halos
of CGCG 97-114 or 97-125, but the disturbed structure of those
objects makes an unambigouous association impossible.  Several
other knots (e.g., K2 a, b in Figure 4, and a string of knots extending
to the northwest of Dw1) appear to be physically related, and
may be parts of tidal features extending from the larger galaxies.
Finally a few objects (a notable example is K1 in Figure~4) appear
to be truly isolated \hii\ regions.  

Photometry of the deep broadband
images reveals a detected stellar component in almost all of these
knots, with typical magnitudes in the range $B \sim R \sim 21 - 23$, 
or absolute magnitudes in the range $-$11 to $-$13, comparable to those
of luminous stellar associations in normal spiral and irregular galaxies
or the faintest dwarf irregular galaxies.   However these knots
are distinguished by their compactness and their strong line emission
(next section), which are more typical of disk \hii\ regions than
of dwarf galaxies.

\subsection{Star Formation Rates and Properties}

The star formation regions are scattered throughout this area
of radius $\sim$50 kpc, most of which are found in galaxies showing 
faint, underlying distribution of stars, and
are very compact.  For example,
Dw1 is comprised of six very compact H$\alpha$ 
regions as seen in Figure~2. Dw3 has a very similar appearance.

The two CGCG galaxies and the three dwarf galaxies are large and
bright enough so that we can measure reliable emission-line fluxes,
equivalent widths (EWs), and star formation rates (SFRs).  These are
listed in Table 3.  Iglesias-Paramo et al. (2002) have also reported 
fluxes and equivalent widths for three of these galaxies (CGCG 97-114,
CGCG 97-125, and Dw1) and the values are in excellent agreement
with our measurements.  The emission-line properties of the two CGCG
galaxies also have been discussed previously by Gavazzi et al. (1998)
and Moss \& Whittle (1998).

As mentioned previously the \halpha\ EWs of CGCG 97-114 and CGCG 97-125
are typical of actively star-forming spiral or Magellanic irregular
galaxies.  The dwarf Dw1 is much more active, with EW(\halpha$+$[NII])
= 101 $\pm$ 4 \AA.  This high EW requires a current SFR that is
several times higher than the time-averaged past SFR in the galaxy,
and is comparable to some of the most extreme starburst galaxies in
the Galactic neighborhood.  Dw3 also has a relatively high SFR.
While Dw1 and Dw3 are characterized by clusters of very compact HII 
regions, Dw2 shows a much more diffuse and fainter H$\alpha$ distribution.  
Its very blue color however suggests that
it is being observed within $10^8$ yr of a recent starburst, or it
is experiencing a short-term lull between bursts.

As alluded to earlier, the isolated \hii\ region knots show much
stronger emission, with EWs in the range 200 $-$ 400 \AA.  The only
local galaxies with such strong line emission are extreme dwarf
starburst galaxies such as I Zw 18 or II Zw 40 (e.g., Martin 1998),
but those galaxies also are at least 10 times more luminous in 
\halpha\ than the knots observed here.  The knots more closely
resemble in EW, luminosity, and continuum color (as best as can
be determined) those of ordinary \hii\ regions in nearby galaxies,
or for that matter the individual \hii\ region knots in the
CGCG galaxies and dwarf galaxies in this region.  Where they differ
is in the lack of an obvious association with a parent galaxy (apart
for the handful of regions in the very outer disks of CGCG 97-114 and
97-125).

The luminosities of the individual \hii\ regions, whether isolated
or within larger parent galaxies, range from a few times $10^{-17}$ 
ergs~cm$^{-2}$~s$^{-1}$ (the effective faint limit of our data)
to $1 - 2 \times 10^{-15}$ ergs~cm$^{-2}$~s$^{-1}$.  The corresponding
\halpha$+$[NII] luminosity range is of order $10^{37} - 10^{39}$
ergs~s$^{-1}$.  This range extends from analogs of large Galactic
\hii\ region complexes (e.g., M17, Carina) to giant \hii\ regions
such as NGC 3603 in the Galaxy or NGC 604 in M33 (e.g., Kennicutt 1984).
Virtually all of the regions are unresolved or semi-resolved in our
images, not suprising given our typical linear resolution of $\sim$500 pc.
Although the distributions of star forming regions are irregular
and/or disturbed in nearly all of the galaxies, the physical properties
of the \hii\ regions themselves do not appear to be unusual in any
discernable way.

\subsection{Radial Velocities}

Table~4 lists the positions and radial velocities of the knots 
that were measured spectroscopically.  
All but one of the detected \hii\ regions have
radial velocities within a narrow range, 8050 $-$ 8300 
km~s$^{-1}$.  These correspond to the velocities of the 
brighter galaxies CGCG~97-114 (8450 km~s$^{-1}$; Falco et al. 1999)
and CGCG~97-125 
(8271 km s$^{-1}$; Haynes et al. 1997).  This confirms the
physical association of the the dwarf galaxies and knots with
this group, and rules out any association with NGC 3860 or the
possibility that these are superimposed objects near the central
velocity of A1367 itself (6600 km~s$^{-1}$).  It is very likely
that most or all of the emission knots in Figures~2 and 4 are also
associated with this system.  Although we do not have spectroscopic
velocities for the other knots, their emission-line flux ratios
in the Ha16 and Ha12 MOSAIC images (with central velocities of
6723 \AA\ and 6683 \AA\ respectively) are similar to those of 
the \hii\ regions in Table~4, and this restricts the velocity
range to approximately 8200 $\pm$ 500 km~s$^{-1}$.

In table~4 we also list the radial velocities measured from the 
WSRT HI line observations, or when the WSRT resolution does not permit
a precise velocity measurement, the range of velocities observed at
the position of the optical pointings. The agreement is in general quite
good and within the mutual uncertainties, further confirming the association
of the emission knots with the galaxies CGCG 97-114 and CGCG 97-125.
In Dw~3, no HI was detected to a 3$\sigma$ limit of 
$1.3 \times 10^8$ M$_{\odot}$.

\subsection{Emission-Line Spectra and Metal Abundances}

For five of the emission regions 
(Dw~1-a, Dw~1-b, Dw~3-a, CGCG 97-114-b, K2-a),
our spectra are of sufficient quality to constrain the emission
properties and metal abundances.  In addition, we obtained
spectra of three \hii\ regions in the southern spiral arm of
CGCG 97-125 for comparison.  We only analyzed spectra with S/N $>$ 10
in all of the diagnostic lines \halpha, H$\beta$, [OII]$\lambda$3727,
[OIII]$\lambda$5007, and [NII]$\lambda$6583.  For most of the
\hii\ regions the Balmer decrement is sufficiently well measured
to provide a robust reddening corrections for the spectra, using
the reddening curve of Cardelli, Clayton, \& Mathis (1989).
However in the faintest objects (K2-a, Dw~3-b), the reddening 
determination was very uncertain, either due to the signal/noise
in the H$\beta$ line or from the presence of strong underlying stellar
H$\beta$ absorption (which we corrected for using a standard absorption
EW of 3 \AA).  In these latter cases we normalized
the forbidden line ratios to \halpha, with an assumed \halpha/H$\beta$
ratio of 3.0.  To confirm the reliability of the data we obtained
spectra of Dw~1-a, b and the CGCG 97-125 \hii\ regions at much 
higher quality with the MMT telescope (\S 2), and these yield consistent
results (see Figure~5).  Nevertheless we will restrict most of our analysis to the
reddening-insensitive [OIII]/H$\beta$ and 
[NII]/\halpha\ ratios, and use empirical abundance indices
(which are more sensitive to the reddening correction) only to
roughly characterize the abundance ranges for the objects.  

The left panel of Figure~6 shows the excitation properties of
the \hii\ regions in the familar plot of [OIII]/H$\beta$ vs
[NII]/\halpha\ introduced by Baldwin, Phillips, \& Terlevich (1981).
The A1367 \hii\ regions are plotted as large symbols, while the
small points represent the samples of nearby \hii\ regions in
spiral galaxies from McCall, Rybski, \& Shields (1985) and 
\hii\ regions in M101 from Kennicutt \& Garnett (1996).  Most of the 
emission regions in this group fall along the standard emission-line 
for stellar-photoionized \hii\ regions.  The main locus of points
appears to lie slightly to the right of the main excitation sequence.
This could be caused by any number of factors, including an enhanced
N/O abundance ratio, contamination of the spectrum by shocked gas
(e.g., Kewley et al. 2001), or by excess stellar absorption at 
H$\beta$, which would artificially elevate the measured 
[OIII]/H$\beta$ ratio.  In the few cases where our spectra extend
to [SII]$\lambda\lambda$6717,6731 we see no evidence of enhanced
[SII], which would be expected if the gas were shocked.  Otherwise
our spectra are not of sufficient quality to distinguish between
those possibilities, but in any case the objects appear to be normal
stellar-ionized \hii\ regions.

We also have used the abundance-sensitive empirial ratios
$R_{23} \equiv$ ([OII]$\lambda$3727 + [OIII]$\lambda$4959,5007)/H$\beta$
(Edmunds \& Pagel 1984) 
and [NII]$\lambda$6583/[OII]$\lambda$3727 (Kewley et al. 2002) to
place crude constraints on the metal abundances in the regions.
The right panel of Figure~6 plots the positions of the A1367 knots
and the comparison \hii\ regions in a diagram with the logarithms
of $R_{23}$ on the horizontal axis and [NII]/[OII] on the vertical 
axis.  As shown in Kennicutt \& Garnett (1996), the two indices form
a relatively tight abundance sequence in normal galaxies, and the
A1367 \hii\ regions appear to fall along the same locus, although
a few objects appear to lie somewhat higher than the mean relation
for spiral galaxies and M101.  A conservative estimate of the 
observational errors is indicated by the spread of values for
CGCG 97-125 (open squares in Figure~6), which include multiple
measurements of some of the same \hii\ regions.  The corresponding
uncertainty in absolute abundances is as much as $\pm$0.2 dex, 
including an uncertainty of $\pm$0.1--0.15 in the zeropoint
of the abundance calibration of the indices in this range
(e.g., Kewley et al. 2002).  Nevertheless this still allows us to
draw some important conclusions about the metal abundances and
physical nature of the emission regions.

The \hii\ regions in this group are surprisingly metal-rich.
If we apply the calibrations of Kewley et al.\ (2002) the range of
$R_{23}$ and [NII]/[OII] values for the \hii\ regions correspond
to oxygen abundances of $12 + \log(O/H) = 8.3 - 8.9$, or 
0.25 $-$ 1.0 (O/H)$_\odot$.  Other calibrations of the line indices 
give a similar abundance range.   The \hii\ regions in the bright
galaxy CGCG 97-125 lie at the high end of this range, at 
solar abundance ($12 + \log(O/H) =  8.9 \pm 0.2$

The most metal-poor regions are those in the isolated dwarf
galaxies Dw~1 and Dw~3 (large open and filled circles in Figure~6,
respectively).  Normal field irregular and spiral galaxies
exhibit a pronounced metallicity-lumninosity relation (e.g., 
Skillman, Kennicutt, \& Hodge 1989, Zaritsky, Kennicutt, \& 
Huchra 1994), and if these dwarf galaxies
have evolved chemically independent of their environment we would
expect their abundances and luminosities to lie on this correlation.
For the Skillman et al. (1989) relation galaxies in this luminosity
range ($M_B = -15.8$ to $-$16.5) have a typical oxygen abundance of
$12 + \log(O/H) = 8.0 \pm 0.3$.  Dw~1 and Dw~3 appear to be more
metal rich than this mean by about a factor of two in $O/H$, but
this difference is marginally significant given the accuracy of 
the abundance measurements, the dispersion in the metallicity-luminosity
relation, and a possible offset between the $R_{23}$ and [NII]/[OII]
index calibrations of Kewley et al.\ (2000) and the electron-temperature
based scale of Skillman et al.\ (1989).

The abundances of the faint isolated knots K2-a and K2-b are completely
anomalous, however.  Their forbidden-line spectra (crosses in Figure~6)
correspond to abundances in the range 0.5 - 1.0 (O/H)$_\odot$.  If these
knots were isolated \hii\ galaxies we would expect their abundances
to be in the range 1/40 -- 1/20 solar!  Instead their metal abundances
are comparable to those of CGCG 97-125 (which is consistent with the
normal metallicity-luminosity relation of Zaritsky et al.\ 1994).
These observed abundances immediately rule out any evolutionary 
scenario in which the intergalactic knots K2 are isolated, normal dwarf
galaxies evolving as closed systems.  The high abundances probably also
rule out a picture in which the K2 knots condensed out of intergalactic
gas in A1367.  Instead, the high metallicities, and their similarity
to the compositions of the \hii\ regions in CGCG 97-125 favor a
picture in which the K2 knots formed from enriched material stripped
from one of the larger galaxies, perhaps in a recent tidal interaction.
The suggestion of a tidal arm of \hii\ regions connecting CGCG 97-125
and CGCG 97-114 mentioned earlier (see Figure~4) would be consistent
with this interpretation.

\subsection{HI Aperture Synthesis Imaging}

The integrated HI column density image of this group of galaxies is
shown in Figure~7 as an overlay on the blue Digital Sky 
Survey.\footnote{The Digitized Sky Surveys were produced at the 
Space Telescope Science Institute under U.S. Government grant 
NAGW-2166. The images of these surveys are based on photographic data 
obtained using the Oschin Schmidt Telescope on
Palomar Mountain and the UK Schmidt Telescope. The plates were 
processed into the present compressed digital form
with the permission of these institutions.}  
CGCG 97-125 is clearly detected and has an HI mass of $3.9 \times
10^9$ M$_{\odot}$. Also CGCG 97-114 (NGC~3860b) is detected, albeit
at a much lower level, and has a measured HI mass of $3.0 \times
10^8$ M$_{\odot}$.  The corresponding M$_{\rm H \rm I}$/L$_{\rm B}$
ratios are 0.31 (0.49 if we include the extended HI features joining Dw 2
and K2 with CGCG 97-125) and 0.03 respectively. These are typical values for
Sc-Sd and E/S0 galaxies, respectively (Roberts \& Haynes
1994).  
This implies that CGCG 97-114 is somewhat gas poor for its Hubble type
(Sa, Haynes et al. 1999) by a factor of $\sim$ 5: only a quarter of the
Sa galaxies have M$_{\rm H \rm I}$/L$_{\rm B}$ values below 0.03
(Roberts \& Haynes 1994). CGCG 97-125 on the other hand has a normal gas
content for its Hubble type (Sc, Haynes et al. 1999).

The relatively high SFR in
CGCG 97-125 is consistent with its large HI content, but the even
higher SFR in CGCG 97-114 is surprising in terms of its low HI mass.
At its current SFR of at least 0.7 $-$ 2 $M_\odot$~yr$^{-1}$ 
(depending on the extinction correction for \halpha, which is 
uncertain), the total HI mass of CGCG 97-114 would be depleted in
$\sim 2 - 5 \times 10^8$ years.  Boselli et al.\ (1997) detected
CO emission in this galaxy, corresponding to a total molecular gas
mass of $4 \times 10^8$ M$_\odot$.  Including this gas doubles
the depletion time, but it still is well under
a Gyr.  This suggests that the galaxy is currently
experiencing an intense, transient burst of star formation.

In addition to the two detected CGCG galaxies there appears to be extended
HI, mostly around CGCG 97-125. The brightest extension is to the
south-west, reaching peak column density levels of $2 \times 10^{20}$
cm$^{-2}$, containing $1.6 \times 10^9$ M$_{\odot}$. On deep
exposures one can see that CGCG 97-125 has optical extensions in this
direction, though not reaching as far as the HI extension. The HI
extension appears to be a continuation of the \halpha\ structure to
the west of CGCG 97-125.  This extended structure is typical of
galactic merger remnants (e.g., Hibbard \& van Gorkom 1997), and 
reinforces the evidence from the optical morphology of CGCG 97-125
for a recent merger event in this galaxy.

To the north-west of CGCG 97-125 another HI extension is found,
reaching column density levels of $10^{20}$ cm$^{-2}$, containing $6.4
\times 10^8$ M$_{\odot}$.  Part of this HI extension coincides with
the position of the dwarf galaxy Dw~2 in Figure~4.  It is notable
that HI gas is {\it not} detected unambiguously in the other dwarfs
Dw~1 or Dw~3, though any emission from Dw~1 might be confused with 
gas in CGCG 97-125 (the WSRT beam is most extended in the N-S direction
connecting the two objects).  If these galaxies had normal HI contents,
we would expect them to have gas masses of order $10^8$ M$_\odot$,
which would be near or just below the $3 \sigma$ detection limit 
of the present observations.

The presence of a prominent HI bridge is the strongest evidence for 
strong tidal interactions in the CGCG~97-114/125 group.
As observed in Figure~7, this bridge extends through many of
the emission regions discussed in this paper, from CGCG~97-125 itself
(and possibly Dw~1) through the bright \hii\ regions to the SW of
the galaxy, the knots K2-a, b, and north through the dwarf galaxy Dw~2.
The HI distribution which is not confined to the vicinity of 
a galaxy as observed here is reminiscent of 
other examples of interacting galaxies, such as the M81-M82-N3077
group (Yun, Ho, \& Lo 1994).  It is significant however that there
is no evidence of a tidal bridge connecting the two brightest
members of the group, CGCG~97-125 and CGCG~97-114.

The HI distribution around CGCG~97-125 is extended not only in the
plane of the sky, but in the velocity dimension as well.
A position-velocity diagram centered on CGCG~97-125 is shown in 
Figure~8.  The velocity distribution shows a regular gradient accross 
the galaxy (the optical major axis of CGCG~97-125 is very close to E/W)
ranging from 8090 up to 8490 km/s. If this reflects the rotation of
the disk, it means a projected rotation speed of 200 km s$^{-1}$,
or $\sim$ 340 km s$^{-1}$ when corrected for an inclination
of 36$^{\circ}$ (Haynes et al. 1999). This is exceptionally high for a
M$_{\rm B}$ = $-$19.1 galaxy, which usually has a rotation speed of 
$\sim$ 200 km s$^{-1}$ (Broeils 1992, Verheijen 1997). We do not have 
sufficient resolution to fully resolve the velocity field of the HI 
surrounding CGCG~97-125, so it is conceivable that some of the gas is
not part of the disk of CGCG~97-125 and has very peculiar velocities.
In summary, both the HI distribution and the HI kinematics of CGCG~97-125 
are quite peculiar.

The velocity structure to the west of CGCG~97-125 reflects the complex
kinematics of the extended HI towards the region K2. This gas blends in 
with the gas in CGCG~97-125 at a velocity of 8230 km s$^{-1}$ and 
shows a regular velocity gradient down to 8060 km s$^{-1}$ at the 
position of K2. The HI to the north-west of CGCG~97-125 (associated with
Dw~2) has a velocity of 8140 km s$^{-1}$.

\section{DISCUSSION}

At the outset of this analysis we considered three physical
explanations for this unusual group of galaxies:  

  1)  The CGCG~97-114/125 group is a normal compact group of
field galaxies observed in projection behind A1367; it is 
nothing more than a usual collection of spiral and irregular
galaxies which happen to be actively forming stars at approximately
the same time.

  2)  The unusual star formation properties of the group are 
the result of a strong interaction with the intergalactic medium
in A1367, caused by shocking of the ISM as these galaxies move
through the IGM with an encounter velocity of 1600 km~s$^{-1}$.

  3)  The unusual star formation properties of the group are
caused by current and/or past tidal encounters between two or more of 
the galaxies in the group, largely independent of the cluster environment
outside of the group.

Our follow-up deep imaging, spectroscopic observations, and HI data
appear to strongly favor the last of these interpretations, but we
first summarize the evidence against the other scenarios.  While it is
certainly true that many normal galaxy groups contain as many as
several strongly star-forming dwarf galaxies, the concentration of so
many starbursting dwarf galaxies and \hii\ galaxies in such a small
region ($\sim$100 kpc) is very unusual, apart from groups containing
strongly interacting galaxies.  We would expect to find far more
quiescent dwarf galaxies in the region if the starbursts we observe
were not triggered by a common physical mechanism.  Moreover, the high
metal abundances of intergalactic knots and the HI tails/bridges
provide direct evidence for the importance of tidal processes.

Induced star formation from IGM interactions is not quite as easily
ruled out, especially because evidence for such processes is found
elsewhere in A1367 (Gavazzi et al. 1995).  However this interpretation
appears to be unlikely on a number of grounds.  None of the galaxies
in the CGCG~97-114/125 group show evidence of the bow-shock structure
in the \halpha\ or broadband images, or asymmetric HI distributions
that is characteristic of the other objects of this type in A1367 and
elsewhere.  And perhaps more significantly, we would expect a 1600
km~s$^{-1}$ encounter between the galaxy ISMs and the A1367 IGM to
produce copious soft X-ray emission, but Chandra maps of this region
do not show evidence of this type of extended emission (Sun et
al. 2001, Sun \& Murray 2002).

Instead, several lines of evidence point to the liklihood that
the unusual star formation properties of this group are triggered by
one or more major tidal interactions within the CGCG~97-114/125 group.
To summarize they include: 1) morphological evidence for disturbed
dynamical structure of CGCG~97-125 and 97-114; 2) presence of massive
tidal structures connecting many of the emission regions in the HI
maps;  3) the low dispersion in radial velocities of the star-forming
galaxies and emission knots, including those located outside of the HI
features; and 4) near-solar metal abundances in some of the apparently
isolated intergalactic \hii\ regions along the HI arm, and
suspiciously high abundances in some of the starbursting dwarf
galaxies.

The most straightforward interpretation of these observations is that
the largest galaxy in the group, CGCG~97-125, has undergone at least
one major tidal encounter with other members of the group, including a
recent merger event that has produced its shell-like outer structure.
These interactions have pulled metal-rich gas out of the galaxy, into
an extended tidal tail or arm, and some of the gas has collapsed to
form \hii\ regions or tidal dwarf galaxies in the HI arms.

Similar structures and star-forming regions are observed in some
nearby examples of interacting galaxy pairs and merger remnants.
Perhaps the closest analog is the Antennae system NGC 4038/9, which
exhibits extended HI arms (Hibbard et al. 2001) with similar HI masses
and a series of
massive star-forming knots that have been proposed to be newly-formed
tidal dwarf galaxies (Mirabel, Dottori, \& Lutz 1992, Braine et
al. 2001).  Tidal dwarf galaxy formation also has been purported to be
occurring in the gaseous arms of other nearby interacting galaxies
(e.g., Duc \& Mirabel 1998, Weilbacher et al. 2000, Braine et
al. 2001).  The morphology of the faint knots observed in the
CGCG~97-114/125 group bear some resemblance to these systems,
particularly in HI, but the main difference is the absence of
continuous stellar counterparts to the HI arms; here the \hii\ regions
are fainter and more isolated.  This might be explained if the tidal
features in this group are older, and gravitationally unbound
from parent galaxies already, or if the efficiency of star
formation in these interactions were much lower for some reason.

The origin of the larger dwarf galaxies Dw~1, Dw~2, and Dw~3 is less
clear.  The unusually blue colors and faint, diffuse underlying stellar
components in these galaxies tempts us to speculate that these objects
too may have been formed relatively recently (e.g., last 1--2 Gyr) in
tidal interactions.  However the observed metal abundances of Dw~1 and
Dw~3 are plausibly consistent with their being old irregular galaxies
that have evolved independently.  
Some qualities that characterize tidal dwarf galaxies include
the lack of dark matter and a small fraction of old stellar population 
(Hunter, Hunsberger \& Roye 2000).  
Deeper imaging (at visible and
near-infrared wavelengths) and measurements of the stellar kinematics,
or HI rotational velocity
would be able to discern the presence of an older stellar population,
if any, 
and test whether these objects contain the dark matter halos expected
for normal dwarf irregular galaxies.  
We may have found an example of a very recent interaction 
in which the disk of one of the galaxies (CGCG~97-114 with its low 
M$_{\rm H \rm I}$/L$_{\rm B}$) was severely disrupted by a much more
massive object (CGCG~97-125), leaving the shreds of the outer disk
behind, which we are now witnessing as small star-forming regions.

It is intriguing to speculate on the eventual fate of the metal-rich
\hii\ regions (e.g., K2).  These objects almost certainly are
newly formed from the tidal debris of the galaxy interactions in
this group, but it is unclear whether the associated star clusters
will remain gravitationally bound to the more massive galaxies or
will form new tidal dwarf galaxies.  Again, more accurate kinematic
observations of the knots and the other galaxies in the region should
allow one to fit a dynamical model to the HI and optical observations,
and constrain not only the orbits of the knots but also the 
mass distributions in the halo of CGCG~97-125 and 97-114. 

We also draw attention to the possible connection between these
types of objects and the isolated compact dwarf galaxies that have
recently been discovered in the Fornax cluster by Drinkwater et al.
(2001) and Phillips et al. (2001).  The latter objects appear to
be either tidally stripped remnants of dwarf galaxies or massive
isolated star clusters.  It is conceivable that some of the 
star-forming regions observed in the CGCG~97-114/125 group may
evolve into isolated intergalactic dwarf galaxies or star clusters
in A1367, though it appears that the precursors to the massive
objects observed in Fornax probably were considerably more massive
than the regions we have observed here.

Finally we remark briefly on the the apparent rarity of groups of
this kind in nearby galaxy clusters.  No other comparable subgroups
or subclusters of star-forming regions have been found elsewhere
in our Abell cluster survey, which covers 25 square degrees and
a search volume of approximately 300 Mpc$^3$ in 8 clusters.
This may not be entirely surprising, because if the star formation
observed in this group has been triggered by tidal encounters it
requires low-velocity interactions of order a few hundred
km~s$^{-1}$ or less, which is much lower than the typical encounter
velocities in these rich Abell clusters.  

These considerations suggest instead that compact groups may
be the most prolific environment for this mode of star and galaxy
formation.  A deep \halpha\ imaging survey of compact groups
by Iglesias-Paramo \& Vilchez (2001) has revealed tidally extended
star-forming regions in 5 of 16 groups surveyed.  These \hii\
regions probably are the closest analogs to the objects studied
in this paper, though most of the emission knots found in the
survey of Iglesias-Paramo \& V\'ilchez (2001) lie on well-defined
tidal arms of large galaxies, or on well-defined tidal bridges
connecting the interacting galaxies.  Perhaps deep imaging of
other compact groups will reveal closer analogs to the 
concentration of star-forming galaxies in the CGCG~97-114/125 group.
Until then this remarkable region appears to be unique.

\acknowledgements

SS acknowledges the support by NASA LTSA program, NAS7-1260.
RCK acknowledges the support of the NSF through grant AST98-11789
and NASA through grant NAG5-8426.  Part of this work was completed
while RCK visited the Kapteyn
Astronomical Institute as Adriaan Blaauw Professor, and the warm 
hospitality of that department is gratefully acknowledged.

\begin{deluxetable}{lc}
\tablecaption{WSRT Observing Parameters}
\tablewidth{0pc}
\tablecolumns{5}
\tablehead{
\colhead{} &
\colhead{} \cr
}
\startdata
Dates of observation           &         13 May 2001, 8 September 2001 \cr
Pointing Center (J2000)        &         11:44:50.0  19:47:0.0 \cr
Central Velocity (heliocentric)&         8300 km s$^{-1}$ \cr
Velocity Range                 &         7220 - 9205 km s$^{-1}$ \cr
Velocity Resolution            &         20.9 km s$^{-1}$ \cr
Sensitivity                    &         0.4 mJy/beam \cr
Resolution (ra x dec)          &         18'' x 50'' \cr
Brightness sensitivity         &         0.25 K \cr
\enddata
\end{deluxetable}

\begin{deluxetable}{lcccc}
\tablecaption{Galaxies in the CGCG~97-113/114/125 Group}
\tablewidth{0pc}
\tablecolumns{5}
\tablehead{
\colhead{Galaxy} &
\colhead{RA } &
\colhead{DEC} &
\colhead{B} &
\colhead{R} \cr
\colhead{} &
\colhead{(J2000)} &
\colhead{(J2000)} &
\colhead{(mag)} &
\colhead{(mag)} \cr
}
\startdata 
CGCG 97-114 & 11:44:47.8 & 19:46:24 & 15.70 $\pm$ 0.03 &   14.92 $\pm$ 0.03   \cr
CGCG 97-125 & 11:44:54.8 & 19:46:35 & 15.37 $\pm$ 0.03 &   14.03 $\pm$ 0.03   \cr
Dw1         & 11:44:54.2 & 19:47:16 & 18.35 $\pm$ 0.05 &   18.30 $\pm$ 0.05   \cr
Dw2         & 11:44:51.3 & 19:47:16 & 18.72 $\pm$ 0.10 &   18.19 $\pm$ 0.10   \cr
Dw3         & 11:44:46.0 & 19:47:40 & 19.09 $\pm$ 0.15 &   18.89 $\pm$ 0.15   \cr
\enddata
\end{deluxetable}


\begin{deluxetable}{lcccc}
\tablecaption{Star Formation Rates of H$\alpha$-emitting galaxies}
\tablewidth{0pc}
\tablecolumns{5}
\tablehead{
\colhead{Galaxy} &
\colhead{Flux} &
\colhead{$L$(H$\alpha$)} &
\colhead{SFR\tablenotemark{1}} &
\colhead{Observed EW} \cr
\colhead{} &
\colhead{(erg cm$^{-2}$ s$^{-1}$)} &
\colhead{($10^{41}$ erg s$^{-1}$)} &
\colhead{((M$_{\odot}$/yr)} &
\colhead{($\AA$)} \cr
}
\startdata
CGCG 97-114  &  6.33 ($\pm 0.06$) $\times$ 10$^{-14}$ & 0.66 & 0.59 &  44 $\pm$ 3 \cr
CGCG 97-125  &  9.51 ($\pm 0.06$) $\times$ 10$^{-14}$ & 0.99 & 0.88 &  29 $\pm$ 3 \cr
Dw1          &  8.67 ($\pm 0.19$) $\times$ 10$^{-15}$ & 0.09 & 0.08 & 101 $\pm$ 4 \cr
Dw2          &  7.81 ($\pm 1.59$) $\times$ 10$^{-16}$ & 0.01 & 0.01 &  16 $\pm$ 4 \cr
Dw3          &  2.88 ($\pm 0.15$) $\times$ 10$^{-15}$ & 0.03 & 0.03 &  45 $\pm$ 3 \cr
\tablenotetext{1}{Star formation rates derived using the calibration of 
Kennicutt (1998), without any correction for extinction in the galaxies.}
\enddata
\end{deluxetable}

\begin{deluxetable}{lccc}
\tablecaption{Spectroscopic Data}
\tablewidth{0pc}
\tablecolumns{4}
\tablehead{
\colhead{Knots} &
\colhead{Position} &
\colhead{Heliocentric Velocity} &
\colhead{HI velocity} \cr  
\colhead{} &
\colhead{(RA \& DEC)} &
\colhead{(km s$^{-1}$)} &
\colhead{(km s$^{-1}$)} \cr
}
\startdata
CGCG 97-114-b &  11:44:46.36 $\:\:$ 19:46:41.4    &   8504 $\pm$ 50 & 8474 \cr
CGCG 97-125-b &  11:44:54.66 $\:\:$ 19:46:11.5    &   8170 $\pm$ 50 & 8250 \cr
Dw~1-a        &  11:44:54.61 $\:\:$ 19:47:32.5    &   8161 $\pm$ 50 & 8140-8230 \cr
Dw~1-b        &  11:44:53.74 $\:\:$ 19:47:31.2    &   8070 $\pm$ 50 & 8140-8230 \cr
Dw~3-a        &  11:44:46.39 $\:\:$ 19:47:40.6    &   8266 $\pm$ 50 &  \cr
Dw~3-b        &  11:44:45.49 $\:\:$ 19:47:45.5    &   8289 $\pm$ 50 &  \cr
K2-a          &  11:44:50.61 $\:\:$ 19:46:02.4    &   8099 $\pm$ 50 & 8091 \cr
K2-b          &  11:44:49.61 $\:\:$ 19:46:04.0    &   8053 $\pm$ 50 & 8091 \cr
\enddata
\end{deluxetable}

\newpage

{\bf Figure Captions:}

\bigskip
{\bf Figure 1:} 
Broadband images of the CGCG 97-113/114/125 group.  At the distance
of 93Mpc, 1~arcmin corresponds to 27Kpc

\smallskip
{\bf Figure 2:}
A continuum-subtracted H$\alpha$ image of the region around NGC~3860
in Abell~1367.  See Figure~4 for identifications of galaxies and emission-line
regions.  Note that the bright object near $K2$ (Fig~4) is a foreground stars

\smallskip
{\bf Figure 3:}
{\it Lower left:} The distribution fo galaxies in Abell~1367 in
the plane of sky.  The position of the CGCG 97-113/114/125 group is indicated
by an open circle.  Also shown are the distributions of A1367 galaxies
in the velocity-declination plane {\it (lower right)} and in the
right ascension-velocity plane.

\smallskip
{\bf Figure 4:}
A H$\alpha$ + [NII] + continuum image. Dwarf galaxies and HII regions
detected in this paper are shown.

\smallskip
{\bf Figure 5:}
Spectrum of Knot~a in the dwarf galaxy Dw~1, obtained with the                               
Blue Channel spectrograph on the 6.5 m MMT telescope. 

\smallskip
{\bf Figure 6:}
Spectroscopic properties of selected HII regions in the CGCG 97-114/125 
group. (Left): Correlation between the reddening-insensitive [OIII]/H$\beta$ and
[NII]/H$\alpha$ excitation indices.  Regions in the group are marked as
large symbols.  The small points are comparision HII regions in nearby
galaxies, as described in the text.  (Right):  Distribution of the
abundance-sensitive R23 and [NII]/[OII] ratios, as discussed in the text.

\smallskip
{\bf Figure 7:}
HI column density distribution in the CGCG~97-114/25 group overlaid
on the XDSS blue image.  Contours are 0.5, 1.0, 2.0, 3.0, 4.0, 5.0
and 6.0 $\times 10^{20}$ cm$^{-2}$.  The angular resolution is 18\arcsec
$\times$ 50\arcsec (R.A. $\times$ Dec).

\smallskip
{\bf Figure 8:}
HI Position-velocity diagram (position angle 90 degrees) centered
on CGCG~97-125.  Contours are $-0.7$, $-0.35$, 0.35, 0.7, 1.4, 2.1, 2.8, 3.5
and 4.2 mJy.  Negative contours are dashed.

\end{document}